\documentclass[doublecol, figures]{epl2}
\usepackage[utf8]{inputenc}
\usepackage{amsmath}
\usepackage{color}
\usepackage{multirow}

\date{\today}

\title{Liquid-solid transitions in the three-body hard-core model}
\author{Tommaso Comparin\thanks{tommaso.comparin@ens.fr} \and
Sebastian C. Kapfer\thanks{sebastian.kapfer@fau.de} \and Werner
Krauth\thanks{werner.krauth@ens.fr}}
\shortauthor{T. Comparin \etal}

\institute{ Laboratoire de Physique Statistique, \'{E}cole Normale
Sup\'{e}rieure/PSL Research University, UPMC, Universit\'{e} Paris
Diderot, CNRS, 24 rue Lhomond, 75005 Paris, France}

\abstract{
We determine the phase diagram for a generalisation of two-and
three-dimensional hard spheres: a classical system with three-body
interactions realised as a hard cut-off on the mean-square distance
for each triplet of particles. Quantum versions of this model are
important in the context of the unitary Bose gas, which is currently
under close theoretical and experimental scrutiny. In two dimensions,
the three-body hard-core model possesses a conventional atomic liquid
phase and a peculiar solid phase formed by dimers. These dimers interact
effectively as hard disks. In three dimensions, the solid phase consists
of isolated atoms that arrange in a simple-hexagonal lattice.  }

\begin{document}

\maketitle

\section{Introduction}
In many fields of physics, interactions between particles
are accurately described through pair potentials. These potentials can
be long-range or short-range, and may combine attractive and repulsive
components, as in the Lennard-Jones interaction.  The minimal model for
pairwise-interacting systems is the hard-sphere potential, which has
played a central role in statistical mechanics and computational physics,
and is also a prominent topic in mathematical research \cite{szasz,
conway, weaire}.

In a number of cases, two-body potentials are insufficient to describe
complex physical behaviour. A notorious example are nuclear forces,
where three-body interactions have long been discussed \cite{glockle1996,
epelbaum2009}.  Explicit three-body terms appear also in spin glasses,
as in the well-known $p$-spin model \cite{gross1984}.

Recently, three-body potentials were studied in cold atomic quantum
gases. One proposed route to enhance their effects consists in
suppressing the pair interactions \cite{petrov2014}.  The unitary Bose gas
at low temperature represents another situation where three-body physics
is relevant.  For some bosonic atoms, the Feshbach-resonance technique
allows to reach the \textit{unitary} regime, where the two-particle
scattering length diverges, and an additional length scale is provided by
a three-body parameter.  Experiments in this regime have lead to direct
evidence of the Efimov effect \cite{efimov1970, kraemer2006, zaccanti2009,
pollack2009}, a spectacular manifestation of three-body physics in the
quantum realm. This has triggered a massive experimental and theoretical effort
on the study of this strongly-interacting system \cite{rem2013, fletcher2013,
makotyn2014, vonstecher2010, yan2014, piatecki2014, rossi2014}.

The above considerations motivate the study of a minimal model of
three-body interactions.  In the current work, we consider a classical
system in which each triplet $(i,j,k)$ of particles interacts through
the potential
\begin{equation}
V_3(R_{ijk}) =
\left\{
\begin{array}{lr}
\infty & \mbox{if } R_{ijk} < R_0 \\
0      & \mbox{if } R_{ijk} > R_0
\end{array}
\right. ,
\label{eq:three_body_hard_core}
\end{equation}
where $R_{ijk}^2=(r_{ij}^2 + r_{jk}^2 + r_{ki}^2)/3$.  The interaction
term of eq.\,\eqref{eq:three_body_hard_core} explicitly appears in
recent theoretical models of the unitary Bose gas \cite{vonstecher2010,
piatecki2014}, where it prevents the collapse \cite{thomas1935} of the
quantum-mechanical wave function.  In the corresponding experimental
system, the three-body parameter $R_0$ is related to the length scale of
the van der Waals interactions \cite{berninger2011, wild2012, wang2012}.

Among the many generalisations of the hard-sphere model (\textit{e.\,g.}
to aspherical and polydisperse objects, or to dimensions different from
three), the three-body hard-core model has not -- to our knowledge --
been studied before.

As for the hard-sphere model, the phase diagram of the three-body hard-core
model is independent of temperature, and transitions are driven purely by
entropy. The crucial distinction of the three-body hard-core model is that two
particles can exist at zero distance from each other, and can bind
into dimers.

We consider the classical three-body hard-core model in two and three
dimensions.
To obtain the infinite-pressure limit, we first solve heuristically the
three-body equivalent of the classic sphere-packing problem\cite{weaire}:
We maximise the packing fraction for several families of structures where
the three-body constraint eq.~\eqref{eq:three_body_hard_core} is tight.
We support our result by simulated-annealing calculations which confirm
our highest-density close-packed structures.  In two dimensions, the
densest structure is a triangular lattice of dimers, while in three
dimensions it is a simple-hexagonal lattice of isolated atoms.  Finally,
we use Monte Carlo (MC) simulations in the NPT ensemble to determine the
phase diagram at finite pressure.  Both in 2D and 3D, we find no other
stable thermodynamic phases besides the liquid and the high-density solid.

\section{Close-packed structures}

\begin{figure}
\centerline{
\includegraphics[width=0.75\linewidth]{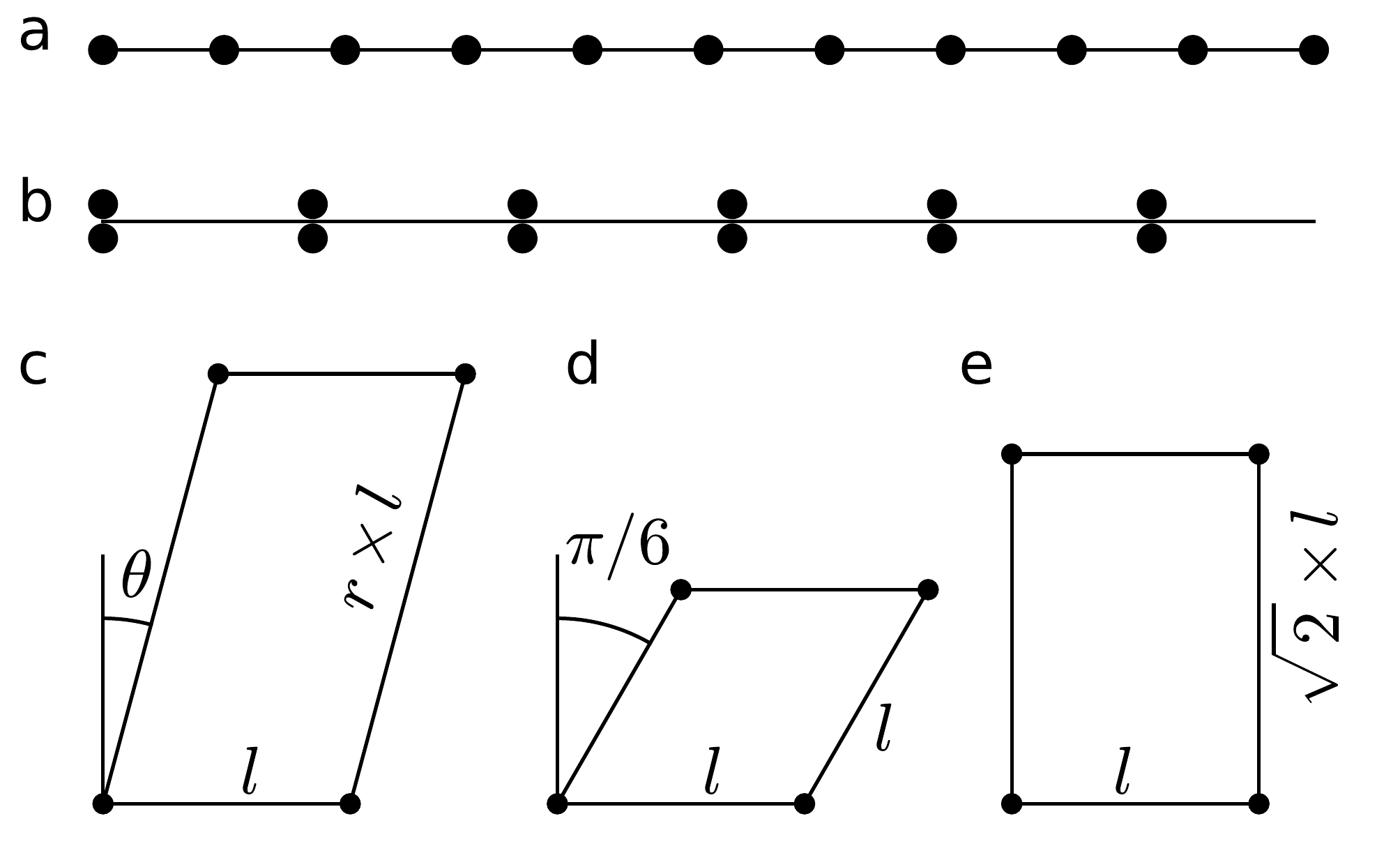}}
\caption{
Close-packed structures in 1D and 2D.
(a)~Linear array of isolated atoms and (b) of dimers with the smallest spacing
allowed by the three-body cut-off.
(c)~General case of the 2D oblique lattice structure,
(d)~triangular lattice at $(r,\theta)=(1,\pi/6)$,
(e)~rectangular lattice with $(r,\theta)=(\sqrt{2},0)$.}
\label{fig:lattices}
\end{figure}

\begin{table}
\caption{
Densities of $D$-dimensional close-packed structures.
The structure may include free parameters (aspect ratio $r$ or angle
$\theta$), and the optimal values are indicated.
Labels \textit{a} and \textit{d} correspond to filling the
lattice sites with atoms or dimers, respectively.}
\label{tab:densities}
\begin{center}
\begin{tabular}{lllll}
$D$ & structure & param. & $\rho_\textrm{max} R_0^D$ &\\
\hline
1
 & regular spacing \emph{a} & -- & $\sqrt{2}$ & $1.41$\\
 & regular spacing \emph{d} & -- & ${2\sqrt{2/3}}$ & $\mathbf{1.63}$ \\
\hline
2
 & oblique \emph{a} & $r=\sqrt{2}$ & $\sqrt{2}$ & $1.41$\\
 & & $\theta=0$ \\
 & oblique \emph{d} & $r=1$ & ${(2/\sqrt{3})^3}$ & $\mathbf{1.54}$\\
 & & $\theta=\pi/6$ \\
\hline
3
 & Barlow ABA \emph{a} & $r=1/2$ & $\frac{13}{27}\sqrt{26/3}$ & $1.42$\\
 & Barlow ABA \emph{d} & $r=\sqrt{2/3}$ & $(2/\sqrt{3})^3$ & $1.54$\\
 & Barlow ABC \emph{a} & $r=1/\sqrt{6}$ & $(2/\sqrt{3})^3$ & $1.54$\\
 & Barlow ABC \emph{d} & $r=\sqrt{2/3} $ & $(2/\sqrt{3})^3$ & $1.54$\\
 & tetragonal \emph{a} & $r=1$ & $(2/\sqrt{3})^3$ & $1.54$\\
 & tetragonal \emph{d} & $r=1$ & $\sqrt{32/27}$ & $1.09$\\
 & simple-hex. \emph{a} & ${r=1/\sqrt{2}}$ & ${2\sqrt{2/3}}$ & $\mathbf{1.63}$\\
 & simple-hex. \emph{d} & $r=1$ & $8\sqrt{2}/9$&$1.26$\\
\end{tabular}
\end{center}
\end{table}

We illustrate our procedure to compute densities of close-packed
structures in the one-dimensional case.  We consider evenly spaced
lattice sites with lattice constant $l$.  For a lattice with a single
atom on every site -- fig.~\ref{fig:lattices}(a) -- the smallest triplet
is composed by three subsequent sites and the hard-core condition in
eq.~\eqref{eq:three_body_hard_core} leads to the following bound on the
lattice spacing:
\begin{equation}
R=\sqrt{\frac{l^2+l^2+(2 l)^2}{3}}=l\sqrt{2} \ge R_0.
\end{equation}
The corresponding upper bound on the density $\rho$ reads $\rho \le
\rho_\mathrm{max}=\sqrt{2}/R_0$.  On the other hand, for the lattice
with every site occupied by a dimer of two atoms, the smallest triplet
consists of a dimer and an additional atom on a neighbouring site; the
root-mean-square distance is $R=l\sqrt{2/3}$ and the close-packed density
is $\smash{\rho_\mathrm{max}=2\sqrt{2/3}/R_0}$.  Thus, at high pressure,
the dimer lattice is favoured (see table~\ref{tab:densities}).

In two dimensions, we consider the family of oblique lattices -- see
fig.~\ref{fig:lattices}(c) -- in which the unit cell is a parallelogram
with edges $l$ and $r\times l$, and  where the smaller internal angle
equals $\pi/2-\theta$.  Without loss of generality, we set $r\ge1$ and
$0\le\theta \le \arcsin(1/(2 r))$.  This family includes the triangular
and the rectangular lattices, as indicated in figs.~\ref{fig:lattices}(d)
and \ref{fig:lattices}(e).  For isolated particles, close packing is
obtained for a rectangular lattice with aspect ratio $\sqrt 2$, while for
dimers, the close-packed structure is a triangular lattice.  Again, high
pressure favours the formation of dimers (see table~\ref{tab:densities}).

In three dimensions, we consider three families of structures: Barlow,
tetragonal, and simple-hexagonal.  Barlow packings are the solutions of
the conventional (two-body) sphere packing problem and include the fcc
and hcp structures  \cite{weaire}.  In our case, the lattice parameters
are regulated by the three-body hard-core repulsion and not by the
hard-sphere diameter.  Within this class, we consider A-B-A and A-B-C
stackings of triangular lattice planes, with a ratio $r$ between the
interplanar spacing and the in-plane lattice constant.  For the tetragonal
structures, we stretch the simple cubic lattice along the $z$ direction
by an aspect ratio $r$.  Finally, we consider the simple-hexagonal
structure, which is an A-A-A stacking of triangular lattice planes.
As for the Barlow structures, the ratio $r$ between the interplanar
and in-plane spacings is a free parameter.  Among the 3D structures
considered, the highest density is achieved by single particles in a
simple-hexagonal lattice with an aspect ratio $r=1/\sqrt{2}$.

The formation of a structure with more than one particle per lattice
site is a peculiarity of systems in which interparticle distances can
vanish \cite{mladek2006}.  In the three-body hard-core model, close-packed
systems with dimers are only favourable in low dimensions, as we now show
by a heuristic argument:  For lattices in which the three closest sites
are all equidistant from each other (examples are the two-dimensional
triangular lattice and the three-dimensional hcp or fcc lattices), we
compute the scaling of the packing density with dimensionality $D$.
We find $\rho_{\max}(\text{dimers}) / \rho_{\max}(\text{atoms}) = 2
(2/3)^{D/2} $.  Thus, a lattice of dimers is favourable only for $D\le3$,
while in higher dimensions the density of the atomic lattice is larger.
(This argument ignores next-nearest neighbours and non-uniform distances
between lattice sites.)  The same scaling argument can be generalised
to a hard-core $k$-body model on a lattice in which the closest $k$
sites are all at the same distance with each other. Up to $(k-1)$
particles can share a lattice site.  As in the $k=3$ case, for large $D$
the largest density is obtained for isolated atoms.

From the list of densities in table~\ref{tab:densities}, we infer that
the close-packed structures -- in the classes considered here -- of the
three-body hard-core model are a regular lattice of dimers in $D=1$, a
triangular lattice of dimers in $D=2$, and a simple-hexagonal lattice of
single particles in $D=3$. Our approach is not exhaustive, and does not
constitute a proof.  In particular, mixtures of atoms and dimers remain to
be explored.  We support our results using a simulated-annealing procedure
\cite{kirkpatrick1983, cerny1985}, in which we perform subsequent short
MC simulations at slowly increasing pressure, starting from low-density
disordered configurations.  Simulated annealing runs for $N=100,72,64$
particles in $D=1,2,3$ dimensions, respectively, indeed yield the expected
structures in the limit of infinite pressure.

\section{Finite-pressure results}  

We now consider the phase diagram of the three-body hard-core model at
finite pressure.  We use a Markov-chain MC scheme to simulate the system
in the NPT ensemble (at fixed particle number $N$ and pressure $P$).
Our MC scheme involves two kinds of moves. The first is the standard local
move, where a randomly chosen particle is displaced in its neighbourhood,
and the move is accepted unless it violates any of the three-body
constraints of eq.~\eqref{eq:three_body_hard_core}.  The second move is a
change of volume that also modifies the shape of the simulation box: For
a $D$-dimensional box with edges $L_1,..,L_D$, the move modifies one of
the $L$'s by adding or removing a random amount of void space (a rectangle
aligned with the coordinate axes, constrained to touch a randomly chosen
particle).  This move does not alter the positions of the particles. As
an example, if particle $k$ and direction $1$ are chosen, one move
corresponds to removing the $D$-dimensional portion of space defined
by $\lbrace(x_1,..,x_D)~|~x_{k,1} < x_1 < x_{k,1}+\delta_1\rbrace$,
where $\delta_1$ is chosen randomly in a fixed interval.  This move,
which changes $L_1$ into $L_1-\delta_1$, is accepted only if it does not
remove any particle and if the new configuration still satisfies  the
three-body constraint.  The complementary move consists in extending $L_1$
to $L_1+\delta_1$, by adding a portion of empty space with edges $\delta_1
\times L_2 \times ..\times L_D$, starting from $x_{k,1}$.  The acceptance
probability for this insertion move is $p_\mathrm{acc}=\exp(-\beta P
\Delta V)$, where $\Delta V=\delta_1 \Pi_{i=2} ^D L_i$ is the change in
volume; this differs from the conventional NPT move, which rescales the
simulation box and the particle positions and requires an additional
factor $V^N$ in the statistical weight.  In addition to the mentioned
acceptance conditions, we reject moves that lead to very elongated box
shapes, $\max_{i,j} (L_i/L_j)>2$, as is common in MC methods based on
variable box shape \cite{filion2009, degraaf2012, bianchi2012} to avoid
strong finite-size effects.

\subsection{Two-dimensional system}

\begin{figure}
\centerline{
\includegraphics[width=\linewidth]{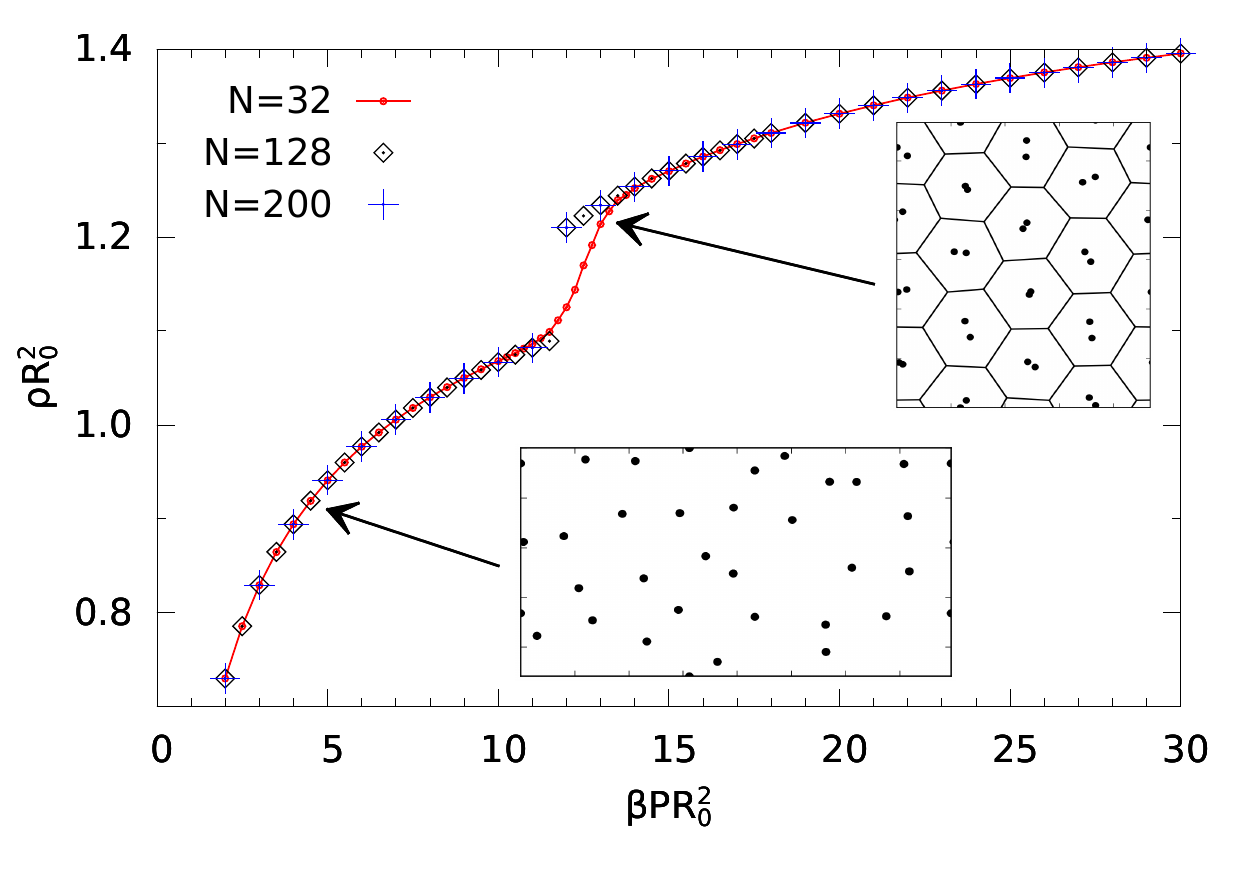}}
\caption{
(Colour on-line) 
Equation of state (density vs.\ pressure) for the 2D three-body hard-core
system, with different system size $N$.
The insets show snapshots of the $N=32$ system at pressures $\beta P R_0^2=4.5$
and $\beta P R_0^2=13$. Black lines in the inset indicate Voronoi cells of the
dimer centres of mass.
\label{fig:2d_eos}}
\end{figure}

\begin{figure}
\centerline{
\includegraphics[width=\linewidth]{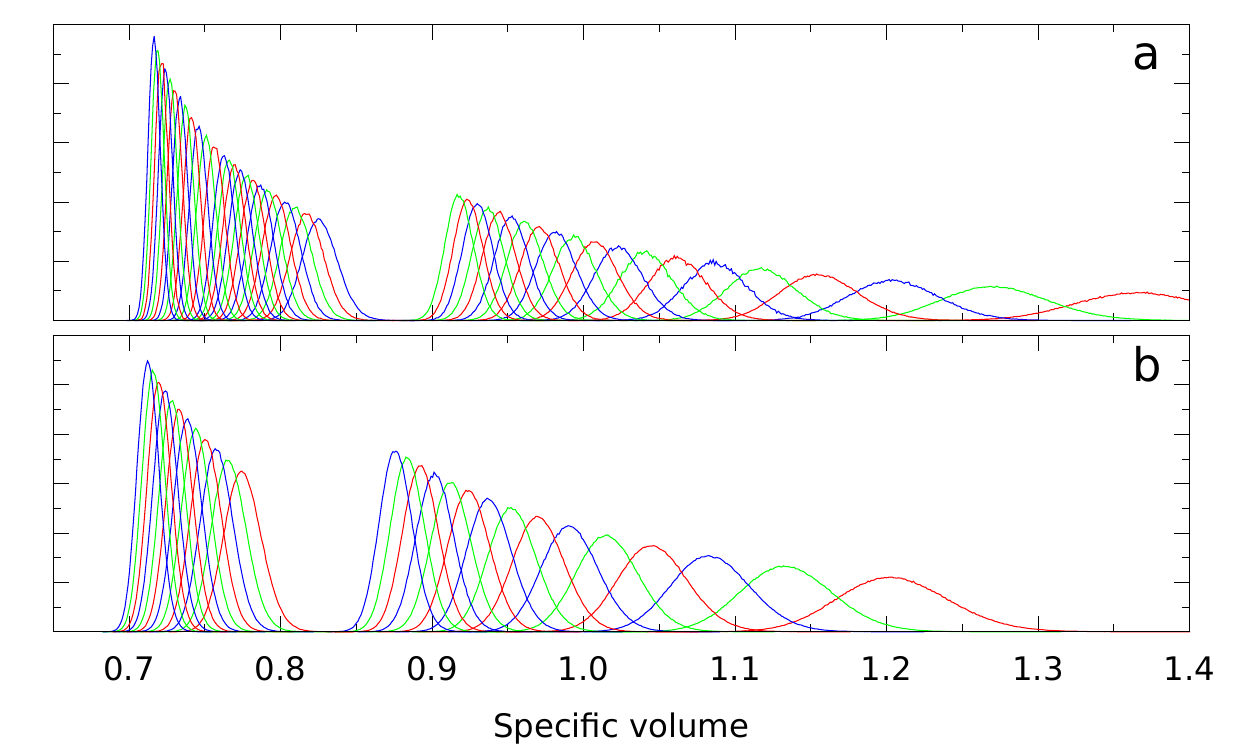}}
\caption{
(Colour on-line) 
Probability density function of the specific volume $V/(N R_0^D)$ at fixed pressure.
(a)~$N=128$ particles in $D=2$ dimensions, (b)~$N=64$ particles in $D=3$
dimensions. Pressures are as in figs.~\ref{fig:2d_eos} and \ref{fig:3d_eos},
respectively.
}
\label{fig:histograms}
\end{figure}

For the $D=2$ system, we expect a disordered isotropic liquid phase at
low pressure and a triangular solid phase of dimers at high pressure.
We initialise the MC sampling with configurations in the solid
phase, and observe melting for low values of $P$. Melting is indeed
observed as a jump in the density, as shown in the equation of state
(fig.~\ref{fig:2d_eos}). For small systems ($N=32$), there is an
intermediate pressure regime where the system oscillates between the
two phases. For larger systems, we only observe a single phase at each
pressure, and the specific volume no longer has a bimodal distribution --
see fig.~\ref{fig:histograms}(a).  The density gap indicates a first-order
phase transition, for the system sizes under study.

In the solid phase, defects are suppressed in our simulations due to
the system sizes considered.  The density in the equation of state,
fig.~\ref{fig:2d_eos}, overestimates infinite-system value (with defects).
In particular, vacancies are not observed on our simulation time scales,
even though our volume- and shape-changing move can in principle create
the void space for a new lattice row to form (which has been advanced
as a mechanism to change the number of lattice sites \cite{swope1992}).

Moreover, in two dimensions, the Mermin-Wagner theorem \cite{mermin1966}
precludes the existence of a crystal with truly long-ranged positional
order, but allows for a solid phase with long-ranged orientational order
and algebraically decaying positional order.  2D systems frequently
exhibit an intermediate hexatic phase with short-ranged positional order,
such as recently shown for hard disks \cite{bernard2011,engel2013} and
soft disks \cite{kapfer2015}.  Since the dimers in our system effectively
interact as hard disks (see below), an intermediate hexatic phase
of dimers might also exist here, and could be observable in larger
simulations.

\begin{figure}
\centerline{
\includegraphics[width=\linewidth]{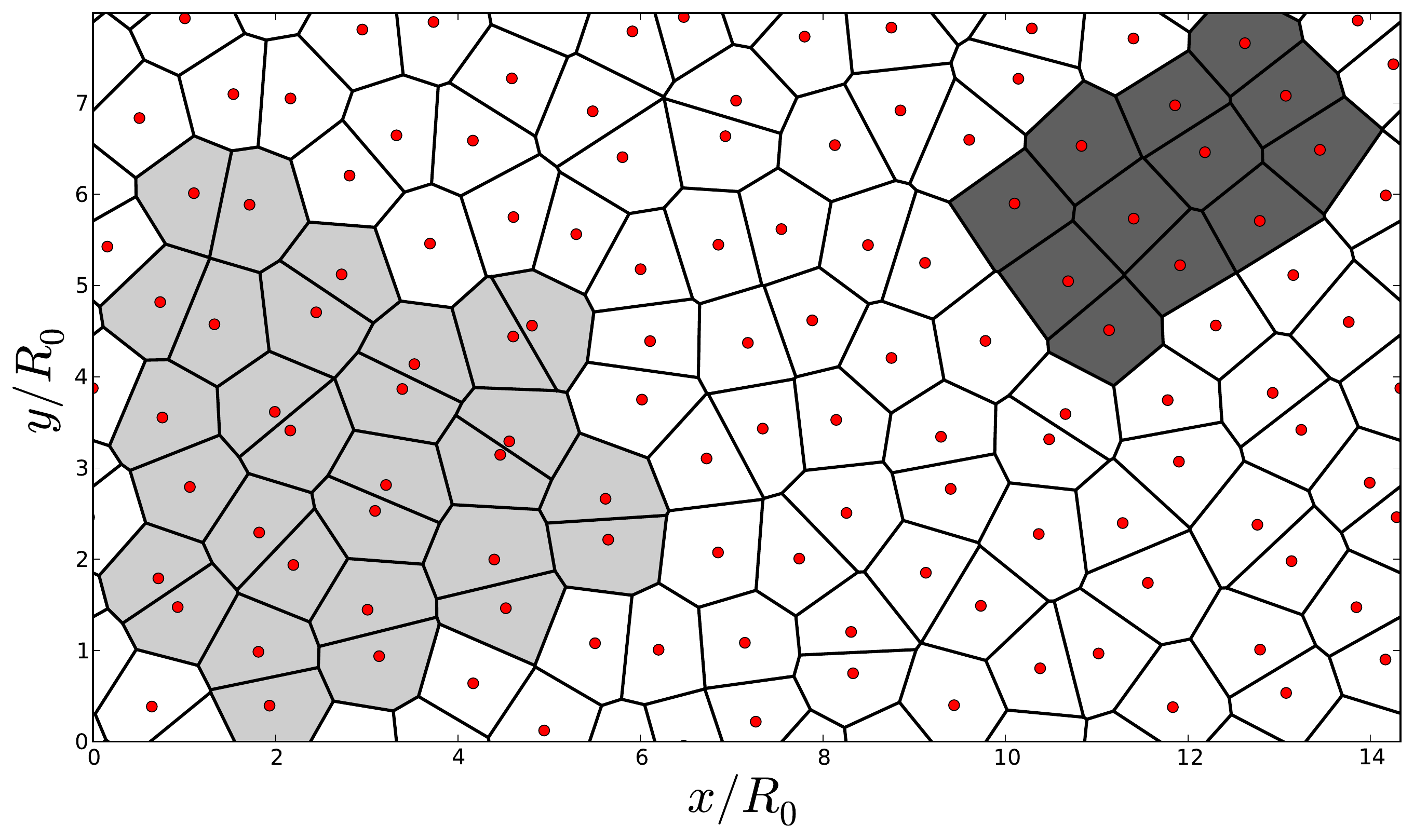}}
\caption{
(Colour on-line) 
Snapshot of $N=128$ particles (red dots) at $\beta P R_0^2=14$, during the
equilibration of a simulation started from a low-density disordered
configuration. Black lines represent the particles' Voronoi cells.
A patch with rectangular order is highlighted in dark gray, while a nucleating 
dimer solid is highlighted in light gray.}
\label{fig:2d_snap_rectangles}
\end{figure}

From table~\ref{tab:densities} we identify a further candidate for
an intermediate phase between the monomer liquid and the dimer solid:
The rectangular close-packed structure of monomers attains a density
only 10\% below the close-packed triangular lattice of dimers.  In our
simulations, we observe transient patches with rectangular order:
fig.~\ref{fig:2d_snap_rectangles} shows both the nucleating dimer
solid and a rectangular patch.  The aspect ratio of the rectangles is
consistent with the close-packed structure, \textit{i.\,e.}, $r=\sqrt 2$.
However, we find no pressure at which the rectangular structure is the
equilibrium phase.  Instead, the rectangular patches seem to decay by
coalescence of two neighbouring lattice rows into a line of dimers,
eventually forming a triangular dimer patch.  Conversely, a dimer liquid
was found to be unstable and to decompose.

\begin{figure}
\centerline{
\includegraphics[width=\linewidth]{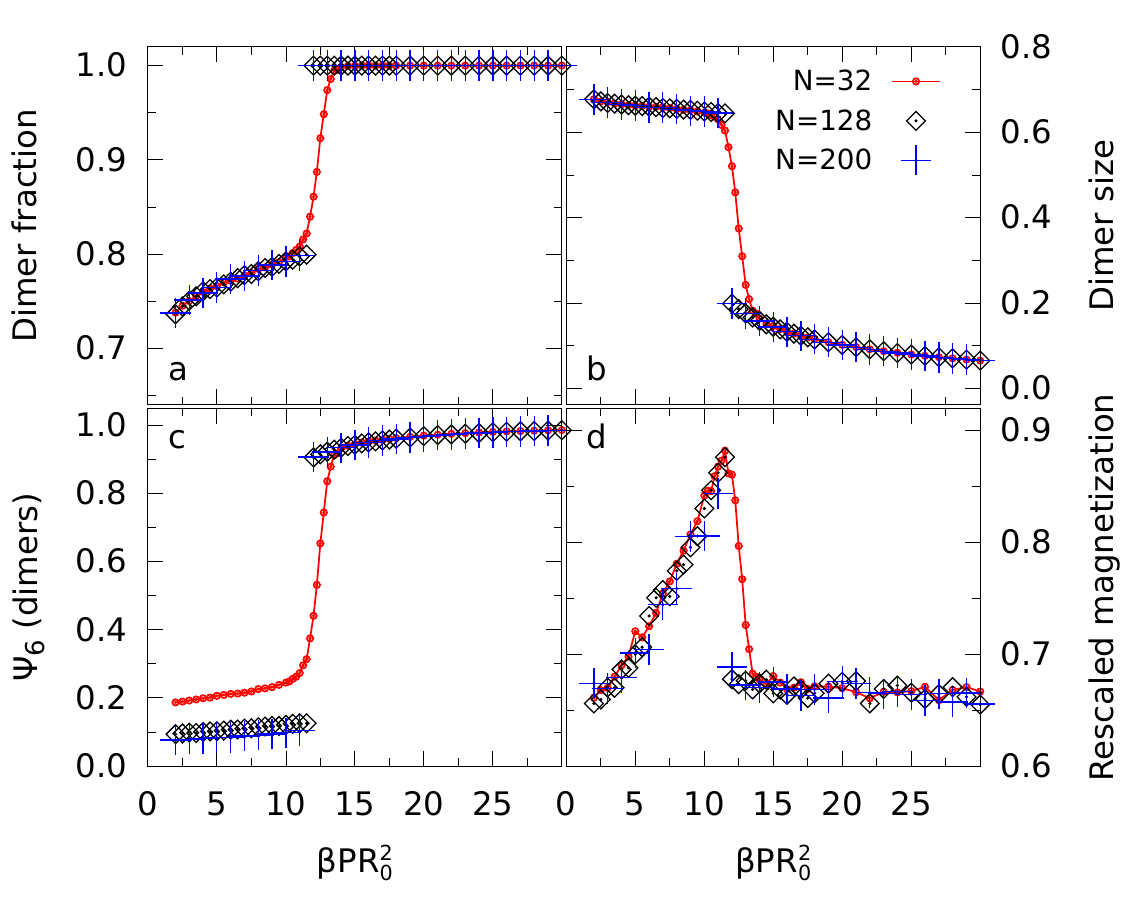}}
\caption{
(Colour on-line) 
Symbols and colours as in fig.~\ref{fig:2d_eos}.
(a)~Fraction $2 N_\text d/N$ of dimers in the system.
(b)~Average dimer size, in units of $R_0$.
(c)~$\Psi_6$ order parameter of dimers.
(d)~Dimer magnetisation $m$, multiplied by $\sqrt{N_\text d}$
(statistical uncertainties are comparable to symbol sizes).
\label{fig:2d_dimer_properties}}
\end{figure}

Quantitatively, we define the term \textit{dimer} to describe two
particles that are each other's closest neighbour. This definition
applies also to the disordered phase, where not all particles participate
in tightly bound pairs.  We measure the dimer fraction $2 N_\text d/N$
according to this definition (where $N_\text d$ is the number of dimers),
and we show in fig.~\ref{fig:2d_dimer_properties}(a) that in the liquid,
this quantity strongly deviates from its ideal-solid value of $1$. Melting
of the solid phase and decomposition of the dimers thus occurs in a
single step.

\begin{figure}
\centerline{
\includegraphics[width=\linewidth]{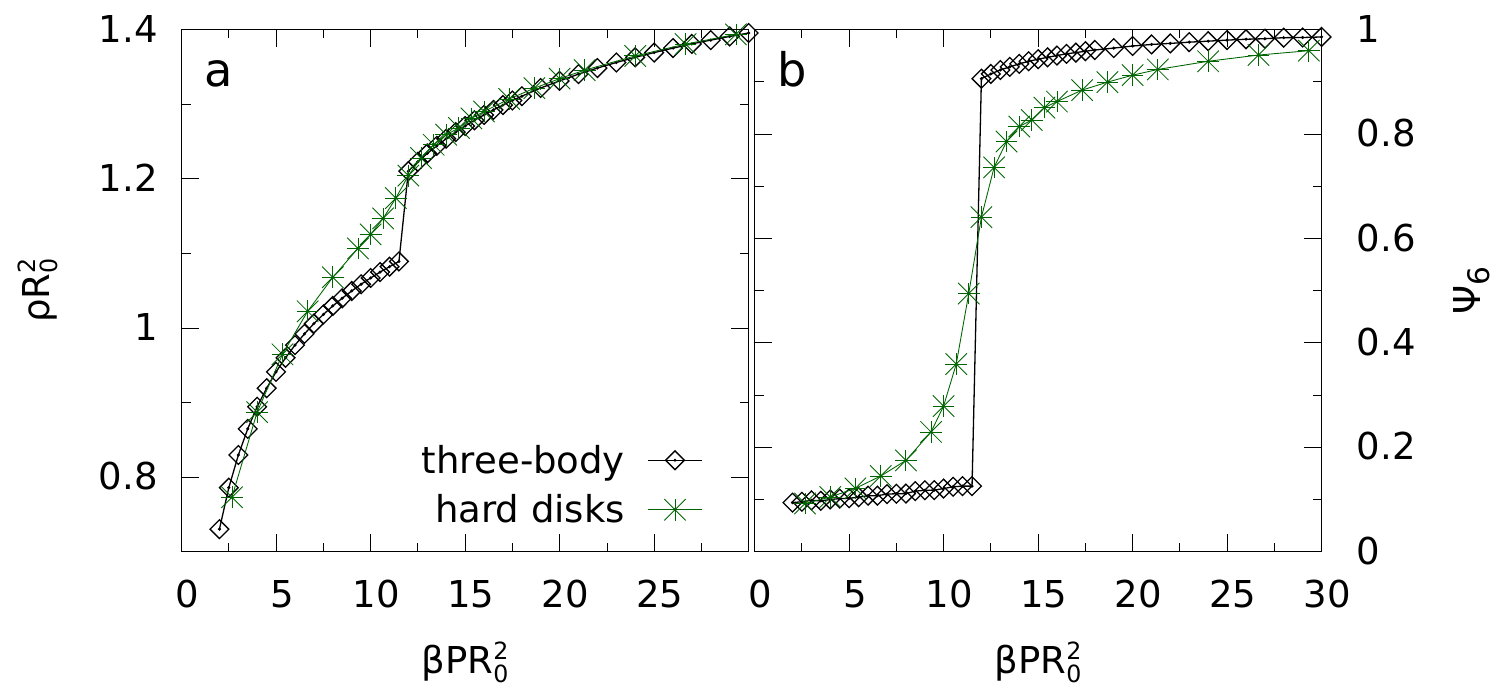}}
\caption{ (Colour on-line) The three-body hard-core solid compared
to hard disks: (a) Equation of state and (b) global orientational
order parameter $\Psi_6$ for $N=128$ three-body-interacting
particles (black diamonds) and $N_\mathrm{disks}=64$ hard-disks
(green stars).  The conversion from hard-disk to three-body units
follows from $\sigma=R_0\sqrt{3/8}$; the density and packing
fraction of the effective model read $\rho_\mathrm{disks}=\rho/2$
and $\eta_\mathrm{disks}=\rho_\mathrm{disks}\pi\sigma^2=\rho R_0^2
3\pi/16$, while the pressure is $P_\mathrm{disks}\sigma^2=2 (8/3)
P R_0^2$.  Data for the three-body model ($\rho$ and $\Psi_6$) are the
same as in figs.~\ref{fig:2d_eos} and \ref{fig:2d_dimer_properties}(c).
} \label{fig:2d_hard_disks} \end{figure}

The three-body model with all particles paired up into small dimers can be
mapped to a conventional two-body hard-core system, containing composite
particles (the dimers).  In the infinite-pressure limit, dimers have zero
extent and the dimer-dimer interaction makes them equivalent to hard disks
with an effective radius $\sigma=R_0\sqrt{3/8}$.  At finite pressure, this
mapping is approximate.  In fig.~\ref{fig:2d_hard_disks}(a), we compare
the equation of state for 128 three-body-interacting particles and 64
hard disks. After rescaling the pressure by a factor of two stemming from
the $2N$-dimensional configuration space of the three-body particles vs.\
the $2N_\text{disks} = N$ dimensions for the hard disks, the equations of
state agree fairly well in the solid phase.  The density of hard disks
is systematically larger, as due to the non-zero extent of the dimers
at finite pressure. This deviation decreases for increasing pressure.

In analogy with the hard-disk model, we define the orientational order
parameter of the liquid--solid transition.  We identify the $N_j$
neighbours of the $j$-th dimer (where $N_j=6$ in the ideal crystal)
through a Voronoi tessellation.  The \textit{local} orientational order
parameter is then \cite{mickel2013}
\begin{equation}
\psi_{6,j}
=
\frac
{\sum_{k=1}^{N_j} W_{jk}\exp(6{\rm i} \theta_{jk})}
{\sum_{k=1}^{N_j} W_{jk}}
,
\end{equation}
where $\theta_{jk}$ is the angle of the vector $\mathbf{r}_j-\mathbf{r}_k$
with respect to a reference axis and $W_{jk}$ is the length of the Voronoi
boundary between dimers $j$ and $k$.  The \textit{global} orientational
order parameter is $\Psi_6=|\sum_{j=1}^{N_\text d} \psi_{6,j} |/N_\text
d$.  In the solid phase this quantity remains finite for increasing system
sizes -- fig.~\ref{fig:2d_dimer_properties}(c) -- while in the fluid phase
it decays as $1/\sqrt{N_\text d}$ and hexagonal order for the residual
dimers is lost.  In fig.~\ref{fig:2d_hard_disks}(b), we show the global
orientational order parameter $\Psi_6$ for the three-body hard-core system
and its hard-disk analogue.  Within our accuracy, the decomposition of
dimers and the melting of the hard disks take place at the same pressure.

At finite pressure, the size of the dimers remains finite.
It is thus interesting to study dimer orientation as an effective
spin model.  Dimers can rotate in 2D and have twofold rotational symmetry.
We define the dimer magnetisation $m$ as
\begin{equation}
m = \frac{1}{N_\text d} \left| {\sum_{j=1}^{N_\text d} \exp(2{\rm i} \alpha_{j})}
\right|
\label{eq:magnetization}
\end{equation}
where $\alpha_{j}$ is the angle that the dimer forms with a reference axis.

\begin{figure}[h!]
\centerline{
\includegraphics[width=\linewidth]{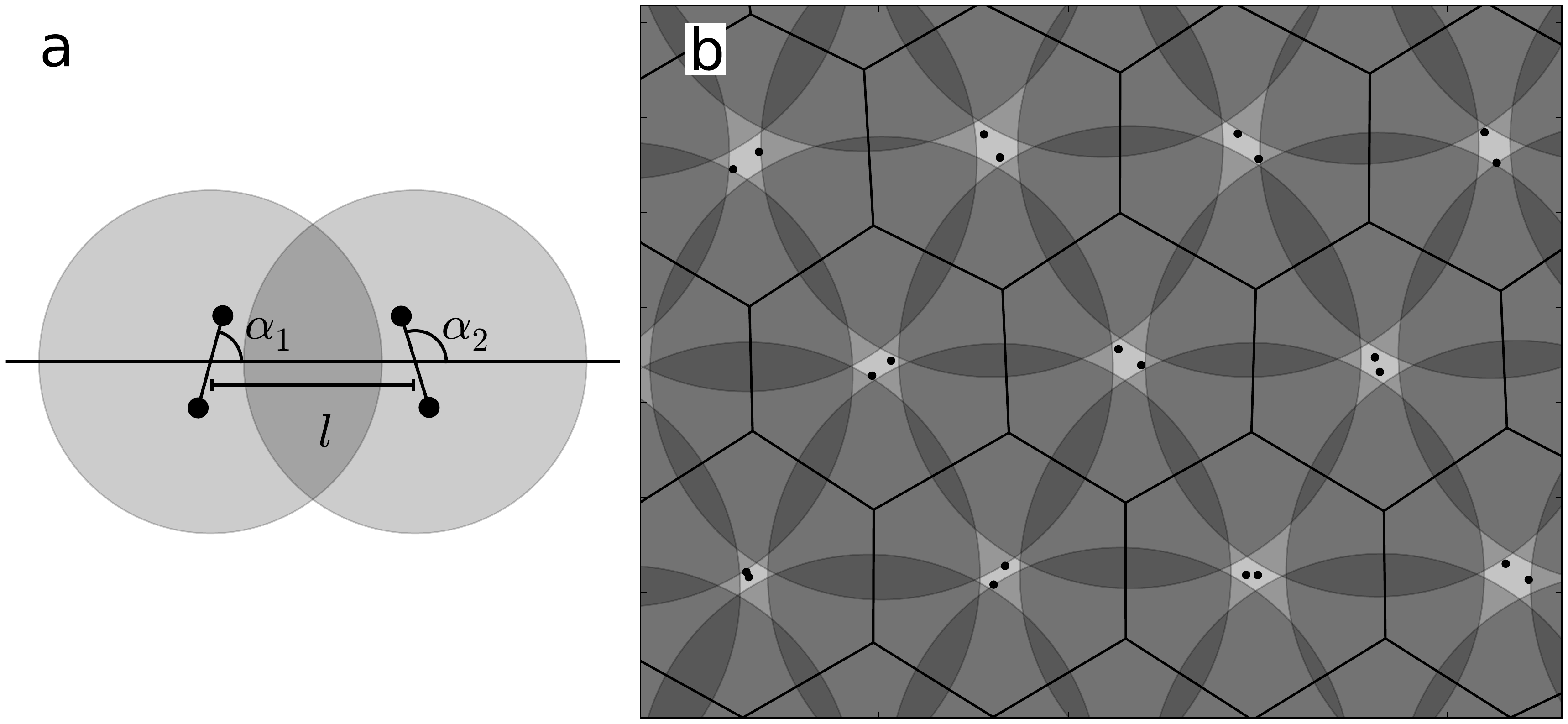}}
\caption{
(a)~Representation of the configuration of four particles (dots)
paired up in two dimers. Each shaded circle represents the excluded region
generated by the two particles in a dimer.
(b)~Partial snapshot of the simulation of $N=128$ particles at
$\beta P R_0^2=20$. For each dimer, the excluded region is shaded. Black lines
indicate Voronoi cells of the dimer centres of mass.}
\label{fig:dimers_excluded_regions}
\end{figure}

We first consider four particles paired up into two dimers of fixed
size, and free to  rotate about their centres of mass, as shown in
fig.~\ref{fig:dimers_excluded_regions}(a).  Each dimer generates
a circular region around its centre of mass into which none of the
two particles of the other dimer can penetrate. The radius of the
excluded region depends on the dimer's size.  When approaching each
other, the dimers can no longer rotate freely and have to progressively
align with each other, standing perpendicular to the dimer--dimer axis
($\alpha_1,\alpha_2=\pi/2$).  In the triangular lattice, this interaction
is frustrated, since not all pairs of dimers in a triangle can align.
The dimers then have to shrink to satisfy the three-body constraint
(see fig.~\ref{fig:dimers_excluded_regions}(b)).  Effectively, the
dimer spins are noninteracting, and their magnetisation vanishes
as $1/\sqrt{N_\text d}$ as the system size increases.  This is
confirmed by the collapse of the $m \sqrt{N_\text d}$ curves -- see
fig.~\ref{fig:2d_dimer_properties}(d).  In the thermodynamic limit,
no magnetic order remains.  In fig.~\ref{fig:2d_dimer_properties}(b) we
show that the average size of dimers in the solid phase keeps decreasing,
as required.

\subsection{Three-dimensional system}
\begin{figure}
\centerline{
\includegraphics[width=\linewidth]{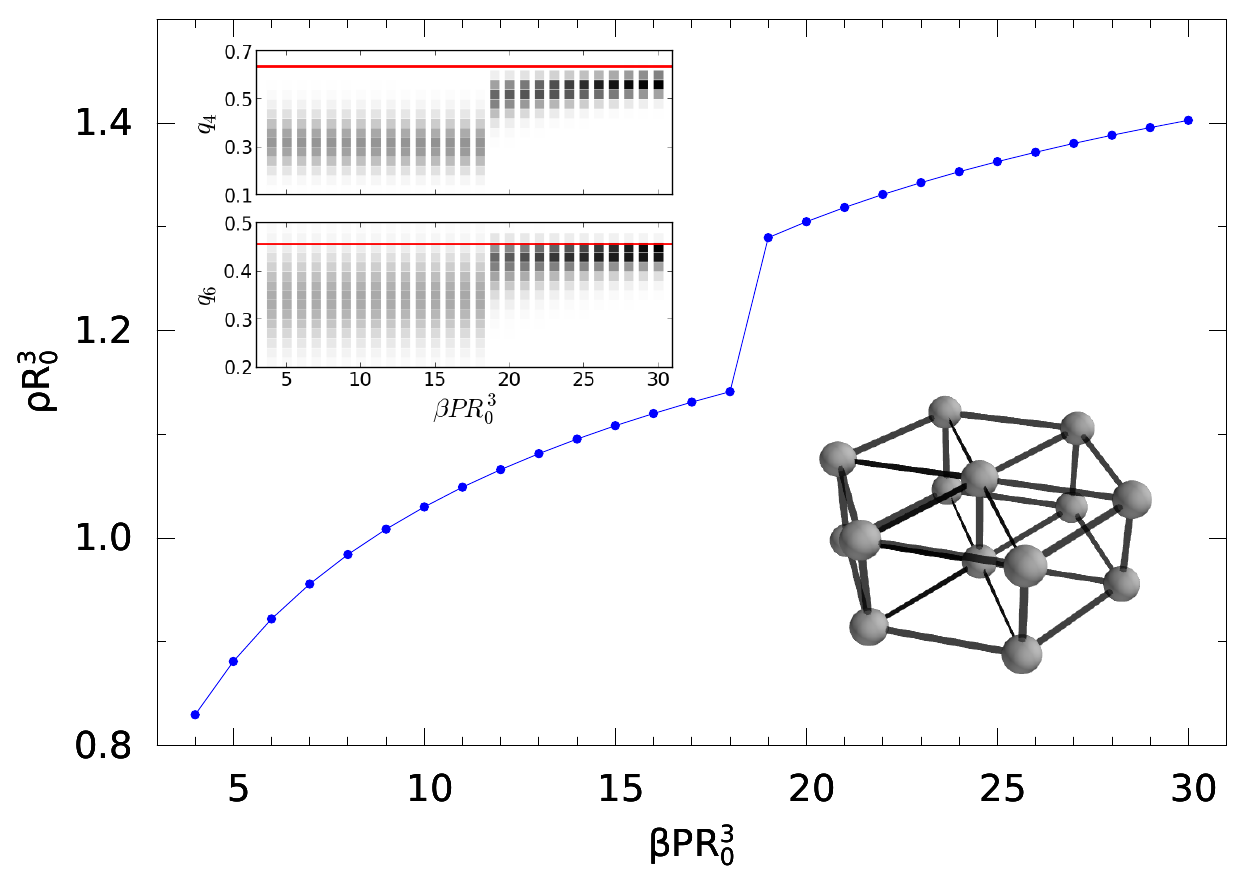}}
\caption{
(Colour on-line) Main panel: Equation of state (density vs.\ pressure)
for the 3D three-body hard-core model of $N=64$ particles.  Inset
(left): $q_4$ and $q_6$ histograms, for each pressure value (darker bins
correspond to higher probability), and perfect-lattice limit (red line).
Inset (right): representation of the simple-hexagonal lattice (realised
at high pressure).  \label{fig:3d_eos}} \end{figure}

Our simulations in three dimensions are analogous to the $D=2$ simulations
described above.  Starting from the ideal simple-hexagonal solid, at low
pressure the system melts into a disordered liquid phase; the equation
of state is shown in fig.~\ref{fig:3d_eos}.  As in the two-dimensional
case, the gap in the specific-volume probability distribution shows
that for this system size, the transition is discontinuous -- see
fig.~\ref{fig:histograms}(b).

We quantify local structure using bond order parameters
\cite{steinhardt1983}, again following the construction of
ref.~\cite{mickel2013} (\textit{i.\,e.}, we weight the contribution of
each bond between neighbours with the area of the shared Voronoi facet).
In the solid phase, the distributions of $q_4$ and $q_6$ (see inset
of fig.~\ref{fig:3d_eos}) are peaked, and their averages approach the
perfect-lattice limit at high pressure.  In the liquid phase, these
distributions are broader, and give no indication for preferred local
configurations.  Except at the melting pressure, the distributions
evolve smoothly with $P$, and there is no sign of further structural
transformations.

\section{Conclusions}

In this paper, we have studied the classical three-body hard-core model
in two and three dimensions, and identified a unique solid phase in
both cases.

In two dimensions, the transition involves simultaneous appearance of
two types of order: particles form dimers, and the dimers order in a
triangular solid that we expect to have quasi-longrange positional order.

The solid phase can be mapped to an effective hard-disk model, which
reproduces the equation of state for large enough pressure.
Dimers break up right at the melting transition of the effective hard-disks.
The small
system sizes considered here do not allow us to comment on the existence
of an intermediate hexatic phase of dimers.  Moreover, we note that the
dimer solid phase does not show (magnetic) order of dimer orientations.
We explain this feature through the frustration of the effective spin-spin
interaction that results from the three-body cut-off.

In three dimensions, the close-packed structure is formed by isolated particles
-- rather than tightly bound dimers -- placed on a simple-hexagonal lattice.
This has an interest in connection with the model for the unitary Bose gas,
since  pairs of particles at very small distances would be pathological in
the original quantum model \cite{piatecki2014}. Our equation of state enriches
the phase diagram in ref.~\cite{piatecki2014} at high temperature and
pressure, and one might be able to link these two different regimes for the
model.

More generally, the nature of the highest-density structure in the
three-body hard-core model might represent a mathematically non-trivial
generalisation of the classic Kepler problem\cite{weaire}, both in two
and in three dimensions.

\section{Acknowledgments} We thank Riccardo Rossi for helpful discussions.

\end{document}